\documentclass[a4paper]{article}

\usepackage{style}      
\usepackage{more}
\usepackage{constants}  

\author[1]{Cong Wang}
\author[2]{Po-Nan Li}
\author[1]{Jana Thayer}
\author[1,*]{Chun Hong Yoon}
\affil[1]{Linac Coherent Light Source, SLAC National Accelerator Laboratory, Menlo Park, CA, USA.}
\affil[2]{Department of Electrical Engineering, Stanford University, Stanford, CA, USA.}
\affil[*]{Corresponding author: {\textnormal{\texttt{yoon82@slac.stanford.edu}}}}

\begin{document}

\title{PeakNet: An Autonomous Bragg Peak Finder with Deep Neural Networks}

\maketitle

\begin{abstract}

Serial crystallography at X-ray free electron laser (XFEL) and synchrotron facilities has experienced tremendous progress in recent times enabling novel scientific investigations into macromolecular structures and molecular processes.  However, these experiments generate a significant amount of data posing computational challenges in data reduction and real-time feedback. Bragg peak finding algorithm is used to identify useful images and also provide real-time feedback about hit-rate and resolution. Shot-to-shot intensity fluctuations and strong background scattering from buffer solution, injection nozzle and other shielding materials make this a time-consuming optimization problem.  Here, we present \peaknet{}, an autonomous Bragg peak finder that utilizes deep neural networks. The development of this system 1) eliminates the need for manual algorithm parameter tuning, 2) reduces false-positive peaks by adjusting to shot-to-shot variations in strong background scattering in real-time, 3) eliminates the laborious task of manually creating bad pixel masks and the need to store these masks per event since these can be regenerated on demand.  \peaknet{} also exhibits exceptional runtime efficiency, processing a $1920 \times 1920$ pixel image around 90 ms on an NVIDIA 1080 Ti GPU, with the potential for further enhancements through parallelized analysis or GPU stream processing.  \peaknet{} is well-suited for expert-level real-time serial crystallography data analysis at high data rates.

\end{abstract}

\section{Introduction}

Serial crystallography with X-ray free electron lasers (XFEL) has enabled radiation damage-free structural determination of macromolecules at room temperature, an approach commonly known as \textit{diffraction-before-destruction} \citep{neutzePotentialBiomolecularImaging2000, chapmanFemtosecondDiffractiveImaging2006,chapmanFemtosecondXrayProtein2011}.  Time-resolved serial femtosecond crystallography at XFEL facilities, like the Linac Coherent Light Source (LCLS), allows the investigation of biochemical reactions with femtosecond-scale time resolution which has led to many significant structural studies \citep{kupitzSerialTimeresolvedCrystallography2014, nangoThreedimensionalMovieStructural2016,pandeFemtosecondStructuralDynamics2016a, youngStructurePhotosystemII2016,sugaLightinducedStructuralChanges2017, kernStructuresIntermediatesKok2018,ibrahimUntanglingSequenceEvents2020, sugaTimeresolvedStudiesMetalloproteins2020} since the first published work at LCLS \citep{aquilaTimeresolvedProteinNanocrystallography2012}.  In the meantime, the data rate at XFEL facilities have also gradually increased from hundreds of Hz to the MHz range, with the European X-ray Free Electron Laser (EuXFEL) being the first MHz facility available to users and LCLS-II set to launch in 2023.  

At the MHz data rate, novel real-time data analysis approaches are needed to tackle two major challenges that arise in SFX experiments, namely data reduction and real time feedback.  The standard approach for data reduction is vetoing events where the x-ray pulse did not ``hit'' the sample. Typically, hit finding in serial crystallography utilizes Bragg peak finding algorithms implemented in such programs as Cheetah  \citep{bartyCheetahSoftwareHighthroughput2014}, DIALS \citep{winterDIALSImplementationEvaluation2018}, and \psocake{} \citep{yoonPsocakeGUIMaking2020}.  Additionally, a deep neural network-based hit finder for serial crystallography has been reported to achieve a processing rate of 1.3 kHz by analyzing downsized 180 $\times$ 180 images on a single GPU \citep{keConvolutionalNeuralNetworkbased2018}.  Real-time feedback in serial crystallography experiments plays a crucial role by providing users with important information about the experimental conditions.  This enables them to make timely decisions that optimize the data collection process. For instance, it is essential to fine-tune the hit rate to maximize the utilization of samples. Additionally, it is important to monitor factors such as the current diffraction limit, crystalline mosaicity, reciprocal space coverage, and other relevant parameters over time. Therefore, relying solely on hit finding is insufficient to gain the necessary insights, making real-time Bragg peak finding vital for delivering prompt and comprehensive feedback.

In this work, we introduce \peaknet{}, a solution that effectively addresses the challenge of autonomously and efficiently identifying Bragg peaks in high data rate scenarios.  Our approach involves transforming the peak finding problem into a semantic segmentation task, where different regions within diffraction images are classified and segmented using a deep neural network. \peaknet{} autonomously detects Bragg peaks while effectively identifying shot-to-shot variations of strong non-Bragg scattering artifacts, eliminating the need for human intervention.  Subsequently, connected areas of pixels are examined and treated as individual peaks, a task easily accomplished through connected component analysis \citep{weaverCentrosymmetricCrossSymmetricMatrices1985}.  The coordinates of the peaks are determined by calculating the center of mass of all pixels belonging to each identified peak.  Notably, the entire peak finding process is performed on GPUs, offering a cost-effective solution for achieving low latency.

\section{Related work}

Reliable and automatic peak finding algorithms have undergone multiple iterations of development.  An early approach involved utilizing template matching \citep{wilkinsonIntegrationSinglecrystalReflections1988a} based on libraries of peak shapes, but this method was found to be slow and inefficient in runtime, especially when dealing with peaks with low signal-to-noise ratios (SNR).  The main obstacle lies in the localization of peaks and the tuning of user parameters for pattern matching, compounded by the difficult task of developing a comprehensive library of peak shapes.  Another influential approach employed region-growing techniques \citep{bolotovskySeedSkewnessMethodIntegration1995, bartyCheetahSoftwareHighthroughput2014} to detect pixel areas with high skewness.  In the context of serial crystallography with XFEL sources, the efficacy of this method was diminished due to the lack of sufficient data to provide reliable statistics, rendering it susceptible to outliers and less adept in analyzing weak peaks at higher scattering angles, which are essential for achieving atomic resolution in structure determination.

While template matching and region-growing techniques require users to provide prior information like threshold values, Robust Peak Finder (RPF) \citep{hadian-jaziPeakfindingAlgorithmBased2017, hadian-jaziDataReductionSerial2021} has demonstrated proficient peak finding capabilities with minimal user input.  This technique utilizes the Modified Selective Statistical Estimator (MSSE) method to segment background pixels and actual peaks.  Furthermore, the RPF method can also parallelize the processing of diffraction patterns, making it ideal for real-time data analysis.

\psocake{} is a software program specifically designed to streamline high throughput data reduction and analysis of XFEL experiments, and is routinely used at SLAC National Accelerator laboratory (SLAC).  The built-in peak-finding algorithm \citep{shinDataAnalysisUsing2018} includes calibrating raw analog digital units (ADUs); applying optional background correction and bad pixel masks; identifying possible peaks; calculating their SNR; determining Bragg spot sizes; and selecting peaks based on size, total intensity and SNR.

In recent years, neural network models have emerged as a promising avenue for Bragg peak finding.  BraggNet \citep{sullivanBraggNetIntegratingBragg2019} is the first method that demonstrates proficiency of neural network models, specifically U-Net \citep{ronnebergerUNetConvolutionalNetworks2015}, in accurately segmenting peak pixels, including weak peaks, from background pixels in neutron crystallography data.  BraggNet employed simulated peaks to create training datasets, which underwent a number of preprocessing steps, including centering and cropping to a specfic size and adding Poisson noise.  It should be noted that BraggNet works on a single peak at a time within a small 32 by 32 window, which deviates significantly from conventional Bragg peak finding tasks that require extracting peak positions from images typically with hundreds or thousands of pixels along one dimension.  Furthermore, another neural network method for Bragg peak analysis was detailed in the work of BraggNN \citep{liuBraggNNFastXray2021}, emphasizing the refinement of Bragg peak positions without the need for explicit fitting of a profile function, such as the pseudo-Voigt profile.  Likewise, BraggNN works on a single peak at a time within a 11 by 11 window, performing regression to two variables representing the peak position.  To the best of our knowledge, no existing neural network based peak-finding models have demonstrated the ability to detect multiple peaks, which is the primary goal we strive to achieve with \peaknet{}.

\section{Methods}

\begin{figure*}[!ht]
\includegraphics[width=\textwidth,keepaspectratio]
{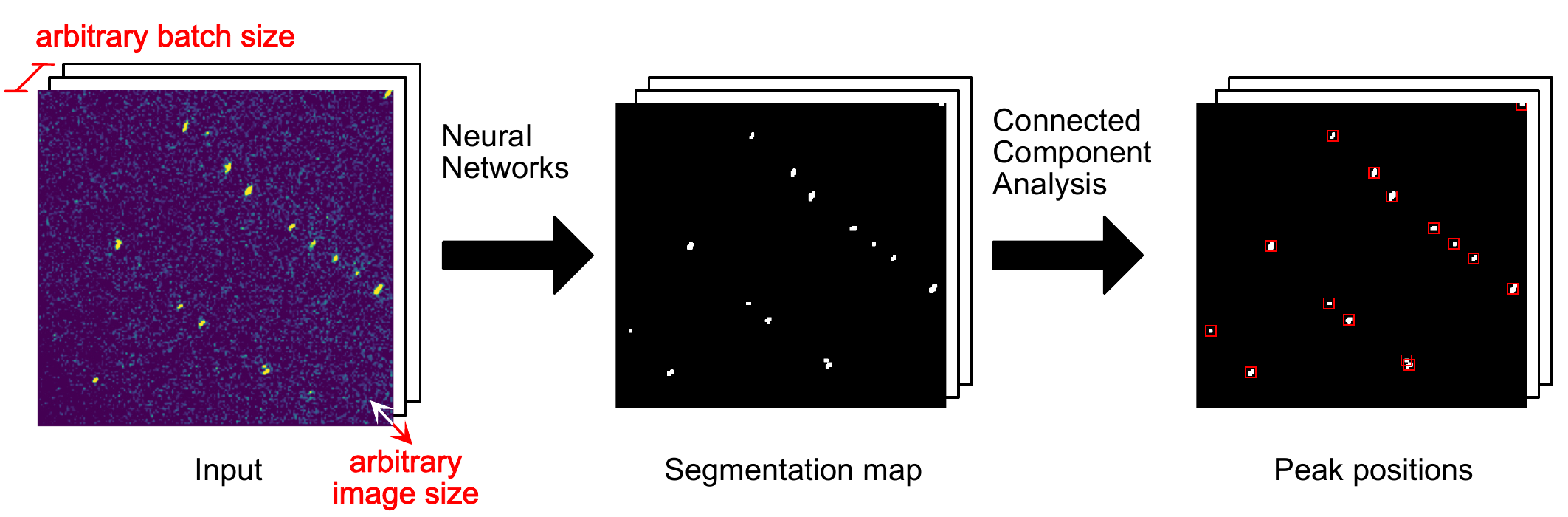}
\caption{The peak finding process of \peaknet{} consists of two steps: (1)
obtaining a segmentation map from the input using the underlying deep neural
network; (2) identifying peak positions based on the generated segmentation map
using connected component analysis.  It is worth noting that \peaknet{} is
capable of accommodating input with arbitrary image sizes and batch sizes.  }
\label{fig : peak finding}
\end{figure*}

The current peak finding process of \peaknet{} comprises of two steps, as depicted in Fig. \ref{fig : peak finding}.  First, a segmentation map is obtained from the input image using a deep neural network.  Subsequently, peak positions are determined based on the generated segmentation map using connected component analysis \citep{weaverCentrosymmetricCrossSymmetricMatrices1985}.  Importantly, \peaknet{} has the capability to process inputs with arbitrary image sizes in batches, enabling efficient analysis.

In this section, our focus is to provide a comprehensive description of the underlying deep neural network components in \peaknet{}.  We will specifically discuss the network architecture, data augmentation techniques, and the loss function employed during the training process.  Furthermore, we will emphasize the use of iterative data curation in addressing the outlier problem.

\subsection{Neural network architecture}

\begin{figure*}[!ht]
\includegraphics[width=\textwidth,keepaspectratio]
{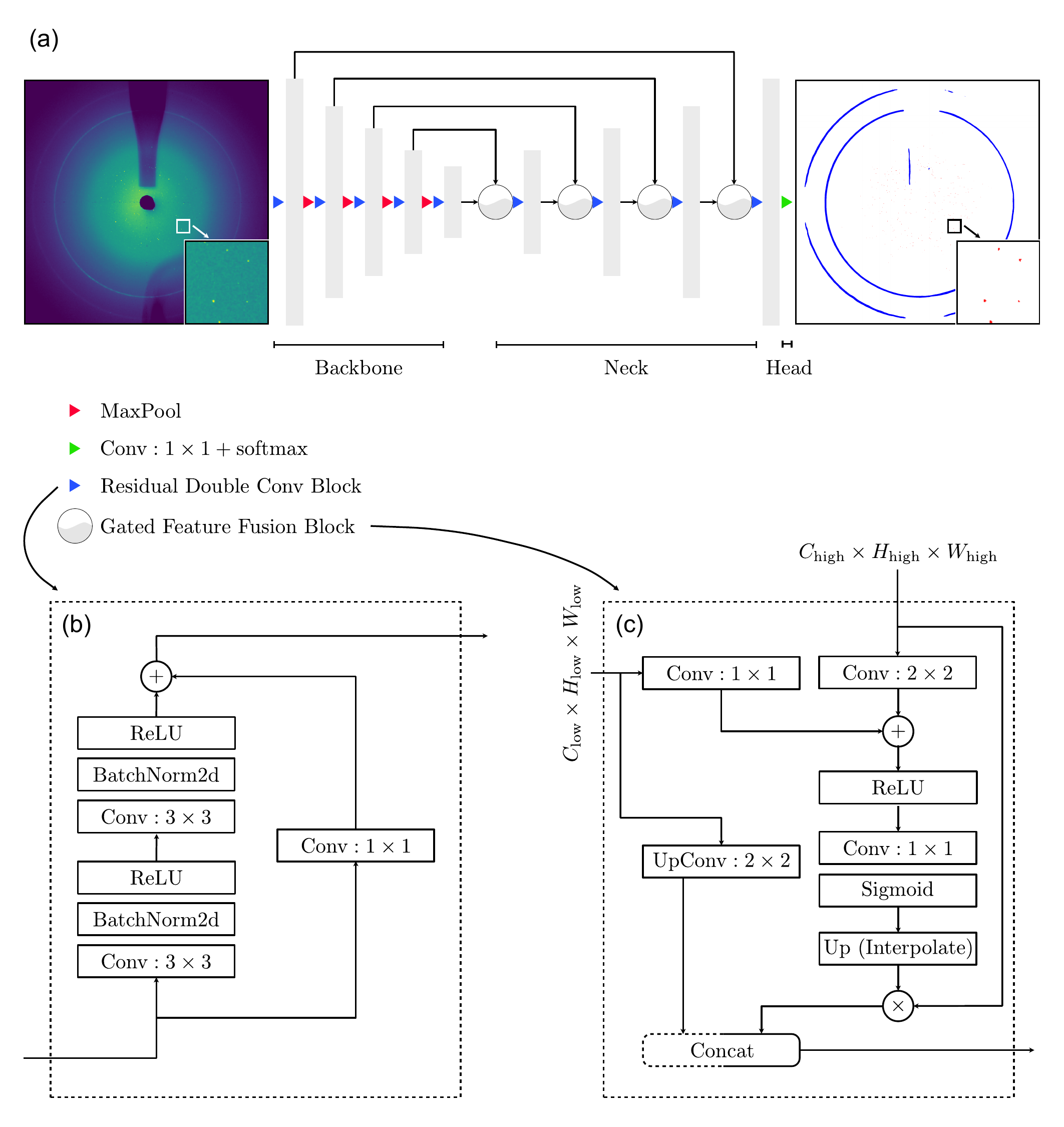}
\caption{Neural network architecture: Attention U-Net with residual connections.
(a) Overall end-to-end network architecture.  (b) Residual double
convolutional block.  (c) Gated feature fusion block for merging
multi-resolution features.}
\label{fig : network arc}
\end{figure*}

The underlying neural network architecture of \peaknet{} is illustrated in Figure \ref{fig : network arc} (a).  It employs a residual attention U-Net architecture, comprising three main components: a feature extraction ``backbone", a feature fusion ``neck", and a single prediction ``head".  The feature extraction ``backbone" utilizes multiple residual double convolutional blocks to extract multi-resolution features, as visualized in Figure \ref{fig : network arc} (b).  In the feature fusion ``neck", low-resolution features are merged with high-resolution features using gated feature fusion blocks, as depicted in \ref{fig : network arc} (c).  Each block incorporates an attention gate that enhances shared features within both low and high-resolution feature maps.  The resulting combined features from each gated feature fusion block are then rearranged by a subsequent residual double convolutional block.  Finally, to accomplish the segmentation task, the prediction ``head" leverages fused features from the ``neck" and generates a prediction map encompassing three distinct labels: background, Bragg peak, and artifact scattering.  The ``head" itself consists of a 1 $\times$ 1 convolutional layer followed by a softmax function.

Our neural network architecture is primarily based on the original ``Attention U-Net" design \citep{oktayAttentionUNetLearning2018}.  However, we want to point out that the three-component design, consisting of the ``backbone", ``neck" and ``head", offers the flexibility to replace these components with different architectures.  For instance, the ``backbone" can be substituted with a ResNet \citep{heDeepResidualLearning2016}, while the ``neck" can be replaced with other multi-resolution feature fusion methods like feature pyramid networks (FPN) \citep{linFeaturePyramidNetworks2017} or bidirectional feature pyramid networks (BiFPN) \citep{tanEfficientDetScalableEfficient2020}.  Additionally, the ``head" can be extended to perform additional tasks, if desired, such as predicting the likelihood of the input image being indexable.  This modular approach provides flexibility and allows for customization based on specific requirements.

\subsection{Focal loss as the loss function}

One main technical issue in training neural networks for Bragg peak segmentation is the extreme peak-background class imbalance (e.g., less than $1 : 10^3$), resulting in reduced prediction accuracy.  This may not pose as a problem in previous peak finders like BraggNet, which operates on a much smaller area of $32 \times 32$, \peaknet{} is trained on larger detectors, such as the Rayonix MX340-XFEL detector with a dimension of 1920 $\times$ 1920 pixels even after $4 \times 4$ pixel binning.  To mitigate the label imbalance, we employ a categorical focal loss to address the problem \citep{linFocalLossDense2018}.

\begin{align}
\text{CFL} &= - \sum_{i = 1}^{N}\sum_{j = 1}^{C} 
            \alpha_j \cdot y_j^{(i)} \cdot (1-\hat{p}_j^{(i)})^\gamma 
            \cdot \log{\hat{p}_j^{(i)}}, \\
\hat{p}_j^{(i)} &= f_\theta(x_j^{(i)})
\end{align}

where CFL stands for categorical focal loss, $\alpha_j$ is a balancing factor for each class, $y_j^{(i)}$ is the ground truth label of the $j$-th class for the $i$-th pixel.  $p_j^{(i)}$ is the predicted probability of the $j$-th class for the $i$-th pixel, calculated by the neural network $f$ parameterized by $\theta$ given input image $x_j^{(i)}$ with $N$ number of pixels and $C$ categories.  $\gamma$ is a parameter controlling the extent to which the weight of well-classified examples, such as background pixels, is reduced.  In our neural network training, we chose $\gamma = 2.0$, essentially, this means $(1-\hat{p}_j^{(i)}) ^\gamma$ is solely responsible for down-weighting easy examples, as $\hat{p}_j^{(i)}$ is usually large in this case.  Finding the ideal values for $\alpha$ can be a delicate process with only marginal returns, as highlighted in the original focal loss study \citep{linFocalLossDense2018}.  In order to fully utilize focal loss, we decided to use a fixed trivial value of $\alpha = 1.2$ for all examples, primarily for the sake of completeness.  This choice essentially treats $\alpha$ as a simple scaling factor without significant impact.  In most cases, the sum of all $\alpha_j$ values, denoted as $\sum_{j = 1}^{C} \alpha_j$, sums to 1.

\subsection{Data augmentation}

\begin{figure*}[!ht]
\includegraphics[width=\textwidth,keepaspectratio]
{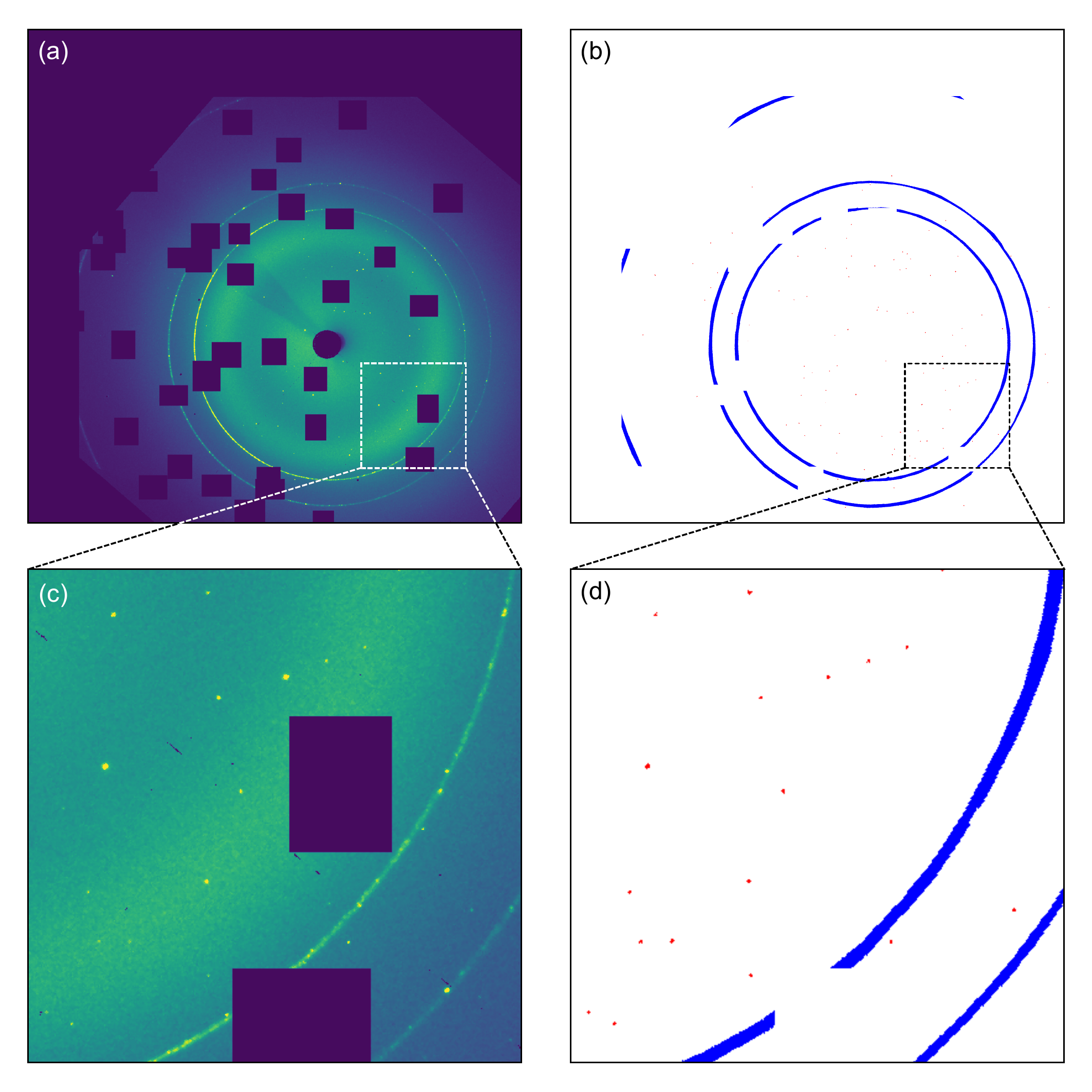}
\caption{Illustration of data augmentation on an input image. (a) Augmented
image with random in-plane rotation, shift, and masking. (b) its corresponding
labeled segmentation map. (c, d) Zoomed-in views of (a) and (b), respectively.
The segmentation label distinguishes red pixels as peaks and blue pixels as
artifact scattering, attributed to scattering from metal shielding and injection
nozzle.}
\label{fig : data aug}
\end{figure*}

We applied three data augmentation techniques to each diffraction pattern in our dataset, including random in-plane rotation, random shifting in both horizontal and vertical directions and random masking.  Random in-plane rotation and shifting augment the training data, reducing the model's tendency to memorize features at specific locations.  This variability promotes more generalized feature extraction and improved generalization performance.  Random masking involves randomly obscuring parts of the input data, forcing the model to learn other relevant features that enhance its predictive capabilities during training. 50 rectangular masks with random sizes between 80 and 120 were positioned randomly. The result of applying data augmentation to an arbitrary example is shown in Fig. \ref{fig : data aug}.

\subsection{Dataset improvement and correction}

\begin{figure}[!ht]
\centering
\includegraphics[width=\columnwidth,keepaspectratio,trim={1in 0in 1in 0in}]
{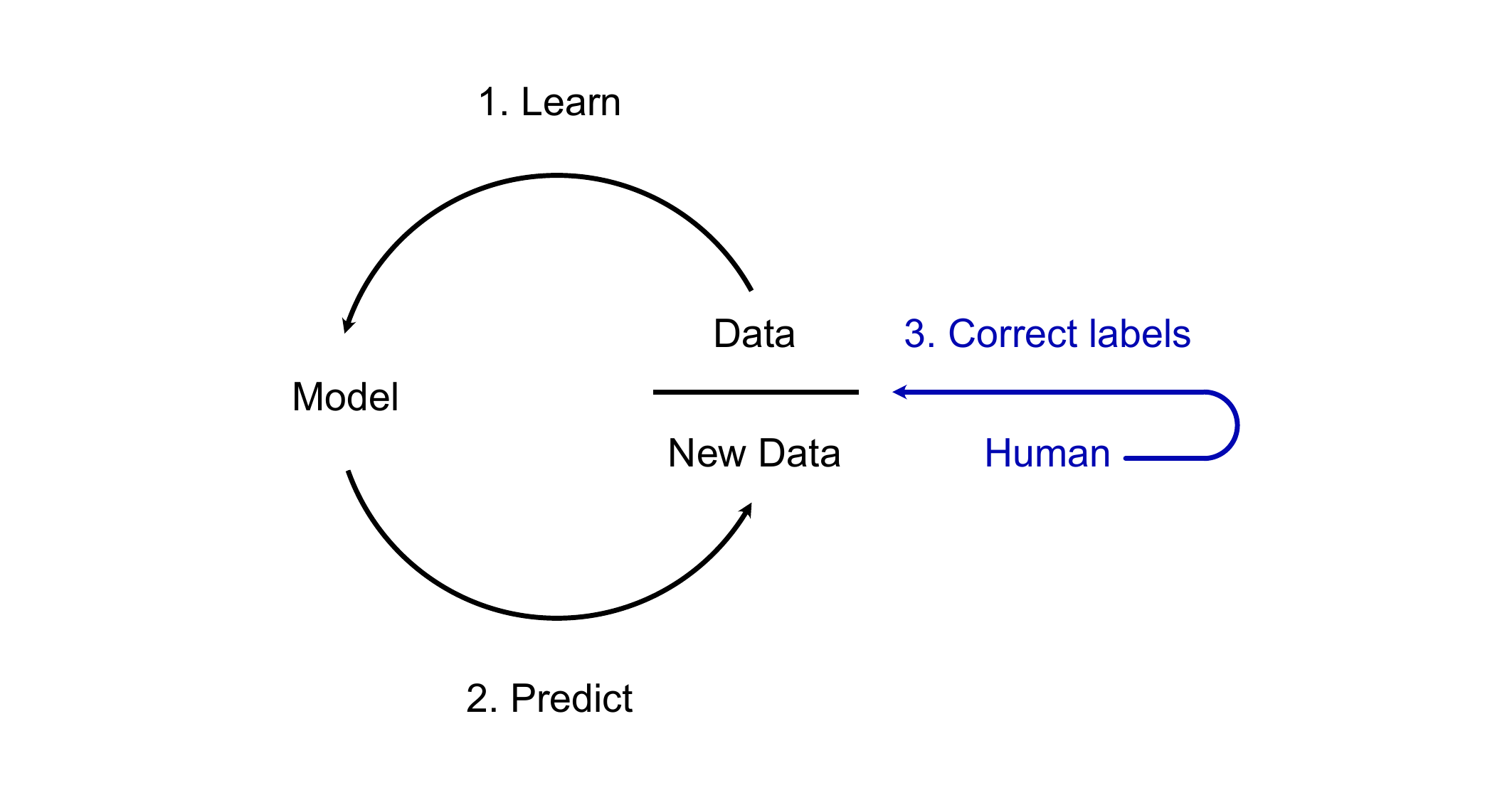}
\caption{The three step process of iterative dataset and model improvement. (1) Model learns from training data. (2) Model predictions reveal potential areas of improvement. (3) Correct low-quality training labels or acquire additional examples of challenging instances.}
\label{fig : data engine}
\end{figure}

\peaknet{} employs a data-driven approach to accurately identify Bragg peaks, requiring dataset improvement and correction to enhance performance.  This contrasts with model-driven approaches that primarily entail direct model modification to improve capabilities.  Fig. \ref{fig : data engine} illustrates a feedback loop for iterative dataset improvement and correction.  

The process begins with the preparation of a source dataset, often accomplished through a combination of algorithmic and manual data labeling.  Subsequently, the deep neural networks within \peaknet{} are trained on this dataset to learn how to accurately detect peaks.  Following training, the model is deployed to predict peaks on previously unseen data, and its performance is thoroughly evaluated.  During this evaluation, instances that notably worsen the performance of \peaknet{} are identified.  To address these limitations, we either correct low-quality labels predicted by the neural networks or, alternatively, acquire additional data that closely resemble these challenging instances and carefully annotate them.  Finally, we merge these newly labeled instances into the training set and retrain the model, thereby improving its overall performance.  

Interestingly, during the initial pass of this iterative process, \peaknet{} showed limited performance when dealing with images containing relatively low levels of photons.  However, by obtaining and annotating these images, we were able to enhance the training dataset and, as a result, significantly improve the predictive performance of \peaknet{}.

\section{Results}

The primary goal of \peaknet{} is to autonomously detect Bragg peaks in diffraction images, devoid of any user input, such as algorithmic parameter tuning or manual artifact masking.  We implemented a two-step approach for peak detection, employing deep neural networks to perform semantic segmentation, which was subsequently followed by connected component analysis to extract image coordinates of Bragg peaks.  In this section, we begin by introducing the datasets used for training, validating, and testing \peaknet{}. Next, we present the evaluation results that demonstrate the predictive performance of \peaknet{}. Finally, we showcase its auto-masking capability and runtime efficiency.

\subsection{Dataset}

We obtained X-ray diffraction images from two experiments conducted at the Macromolecular Femtosecond Crystallography (MFX) instrument within LCLS.  The primary dataset utilized for training and evaluating \peaknet{} is ``mfx13016" (Run number: 28,31-34, 36-38), which contains diverse biological samples, e.g.  thaumatin and proteinase, and this experiment was originally used for automated drop dispensing research \citep{suSerialCrystallographyUsing2021}.  These diffraction images were recorded on a Rayonix MX340-XFEL detector with a dimension of 1920 $\times$ 1920, downsized by a factor of 4 to accommodate a higher readout rate.  

The dataset ``mfx13016" exhibits a notable ring-like background scattering accompanied by bright Bragg peak-like spots located on the rings.  This characteristic presents two primary challenges for crystallography data analysis: increased difficulty in crystal indexing due to false-positive peaks and inaccurate integration results caused by artifact scattering.  The current best practice involves manual masking of these artifact scattering, enabling data analysis to proceed as if they are absent, but this process is time-consuming and can not adjust to shot-to-shot variations in the artifacts in real-time.  Additionally, the final outcomes often hinge on the expertise of the individuals conducting the data processing, resulting in biases, scalability limitations, lack of standardization, and the possibility of human errors.  Nevertheless, the inherent challenges presented by this dataset offer an exceptional opportunity to showcase the autonomous peak finding capability of \peaknet{}.

The neural network training dataset consists of 98 Rayonix images in run 28 of ``mfx13016", which includes thaumatin samples, and an additional 37 images sampled from an internal LCLS experimental dataset related to therapeutics development associated with SARS-CoV-2.  Incorporating the SARS-CoV-2 dataset aims to educate \peaknet{} about the presence of images without artifact rings, diversifying its understanding of different data characteristics.  Nonetheless, all these images used for neural network training have a dimension of 1920 $\times$ 1920 and they are divided into an 80\% training set and a 20\% validation set.  The final test dataset for evaluating the model comprises all images in run 31-34 and run 36-38 of ``mfx13016".

\subsection{Evaluating \peaknet{}'s predictive performance}

The peak finding performance of peak finders is evaluated through three key indicators from a user's standpoint: the number of successfully indexed hits, the indexing rate, and the merging statistics.  For our evaluation, we utilized the CrystFEL software suite \citep{whiteCrystFELSoftwareSuite2012} to perform crystal indexing, integration, and merging tasks.  We use \psocake{}'s peak finder as a baseline in this evaluation.

\subsubsection{Indexing results}

\peaknet{} and manually fine-tuned \psocake{} exhibit comparable peak finding performance in terms of the number of hits and the number of indexed images, as shown in Table \ref{tb : index}.  However, we emphasize that while \peaknet{} operates entirely autonomously, \psocake{}'s performance is contingent upon manual parameter fine-tuning and manual artifact scattering masking.  For example, an earlier work \citep{suSerialCrystallographyUsing2021}, which processed the same experimental data, reported significantly lower hit count and indexed image count.  Consequently, this highlights \peaknet{}'s advantage in user-independent expert-level data processing.

Moreover, while users might prioritize the absolute number of indexed images as a more straightforward metric, the indexing rate provides valuable insights into the method's efficiency.  Both metrics jointly offer a more comprehensive evaluation of the peak finding method. In the case of \peaknet{}, both consistently high number of indexed images and high indexing rate testify to its effective and reliable autonomous peak finding process.

\begin{table*}

\caption{Comparison of crystal indexing results for different methods on the
test dataset.  The table shows the number of hits, number of indexed hits, and
indexing rates for \peaknet{}, \psocake{} (fine-tuned) and \psocake{} (Su 2021).
The ``Run (Samples)" column lists the specific run numbers and biological
samples (TH: thaumatin; PK: proteinase).}

\label{tb : index}

\resizebox{1.0\textwidth}{!}{
\begin{tabular}{lllllllll}
\hline
\multicolumn{1}{l}{Methods}                                                                            & \multicolumn{1}{l}{Run (Samples)}        & \multicolumn{1}{l}{31 (TH)} & \multicolumn{1}{l}{32 (TH)} & \multicolumn{1}{l}{33 (TH)} & \multicolumn{1}{l}{34 (TH)} & \multicolumn{1}{l}{36 (TH)} & \multicolumn{1}{l}{37 (PK)} & 38 (PK) \\ \hline
\multicolumn{1}{l}{\multirow{3}{*}{\peaknet{}}}                                                        & \multicolumn{1}{l}{Number of hits}       & \multicolumn{1}{l}{\textbf{1205}}    & \multicolumn{1}{l}{\textbf{3248}}    & \multicolumn{1}{l}{\textbf{5797}}    & \multicolumn{1}{l}{\textbf{3949}}    & \multicolumn{1}{l}{\textbf{1184}}    & \multicolumn{1}{l}{1030}    & \textbf{1223}    \\ \cline{2-9}
\multicolumn{1}{l}{}                                                                                   & \multicolumn{1}{l}{Number of indexed}    & \multicolumn{1}{l}{\textbf{860}}     & \multicolumn{1}{l}{\textbf{2856}}    & \multicolumn{1}{l}{\textbf{3433}}    & \multicolumn{1}{l}{\textbf{2401}}    & \multicolumn{1}{l}{\textbf{919}}     & \multicolumn{1}{l}{953}     & 1169    \\ \cline{2-9}
\multicolumn{1}{l}{}                                                                                   & \multicolumn{1}{l}{Indexing rate}        & \multicolumn{1}{l}{\textbf{0.71}}    & \multicolumn{1}{l}{\textbf{0.88}}    & \multicolumn{1}{l}{\textbf{0.59}}    & \multicolumn{1}{l}{\textbf{0.61}}    & \multicolumn{1}{l}{0.78}    & \multicolumn{1}{l}{0.93}    & 0.96    \\ \hline
\multicolumn{9}{l}{}                                                                                                                                                                                                                                                                                                                                \\ \hline
\multicolumn{1}{l}{\multirow{3}{*}{\begin{tabular}[c]{@{}l@{}}\psocake{}\\ (fine-tuned)\end{tabular}}} & \multicolumn{1}{l}{Number of hits}       & \multicolumn{1}{l}{1193}    & \multicolumn{1}{l}{3087}    & \multicolumn{1}{l}{5769}    & \multicolumn{1}{l}{3926}    & \multicolumn{1}{l}{917}     & \multicolumn{1}{l}{\textbf{1070}}    & 1212    \\ \cline{2-9}
\multicolumn{1}{l}{}                                                                                   & \multicolumn{1}{l}{Number of indexed}    & \multicolumn{1}{l}{849}     & \multicolumn{1}{l}{2657}    & \multicolumn{1}{l}{3375}    & \multicolumn{1}{l}{2364}    & \multicolumn{1}{l}{760}     & \multicolumn{1}{l}{\textbf{1005}}    & \textbf{1202}    \\ \cline{2-9}
\multicolumn{1}{l}{}                                                                                   & \multicolumn{1}{l}{Indexing rate}        & \multicolumn{1}{l}{\textbf{0.71}}    & \multicolumn{1}{l}{0.86}    & \multicolumn{1}{l}{\textbf{0.59}}    & \multicolumn{1}{l}{0.60}     & \multicolumn{1}{l}{\textbf{0.83}}    & \multicolumn{1}{l}{\textbf{0.94}}    & \textbf{0.99}    \\ \hline
\multicolumn{9}{l}{}                                                                                                                                                                                                                                                                                                                                \\ \hline
\multicolumn{1}{l}{\multirow{3}{*}{\begin{tabular}[c]{@{}l@{}}\psocake{}\\ (Su2021) \end{tabular}}}    & \multicolumn{1}{l}{Number of hits}       & \multicolumn{1}{l}{1083}    & \multicolumn{1}{l}{2454}    & \multicolumn{1}{l}{5604}    & \multicolumn{1}{l}{3756}    & \multicolumn{1}{l}{363}     & \multicolumn{1}{l}{683}     & 771     \\ \cline{2-9}
\multicolumn{1}{l}{}                                                                                   & \multicolumn{1}{l}{Number of indexed}    & \multicolumn{1}{l}{271}     & \multicolumn{1}{l}{1094}    & \multicolumn{1}{l}{963}     & \multicolumn{1}{l}{770}     & \multicolumn{1}{l}{266}     & \multicolumn{1}{l}{569}     & 709     \\ \cline{2-9}
\multicolumn{1}{l}{}                                                                                   & \multicolumn{1}{l}{Indexing rate}        & \multicolumn{1}{l}{0.25}    & \multicolumn{1}{l}{0.45}    & \multicolumn{1}{l}{0.17}    & \multicolumn{1}{l}{0.21}    & \multicolumn{1}{l}{0.73}    & \multicolumn{1}{l}{0.83}    & 0.92    \\ \hline
\end{tabular}
}

\end{table*}

\subsubsection{Merging statistics}

Merging statistics confirms \peaknet{}'s robust performance in terms of consistency and quality in data processing.  For the thaumatin (TH) sample, \peaknet{} demonstrates marginally superior performance in terms of R-split values, correlation coefficients, $I/\sigma(I)$ values, redundancy, and completeness, achieving these results autonomously.  Notably, the R-split value for thaumatin is slightly lower for \peaknet{} (0.2605 vs. 0.2732), suggesting slightly better internal consistency. Similarly, correlation coefficients for \peaknet{} are marginally higher (CC$_1/2$: 0.9434 vs.  0.9368; CC*: 0.9853 vs. 0.9835), indicating a minor advantage in the precision of measurements.

However, for the proteinase (PK) sample, the performance comparison is mixed.  Although \peaknet{} lags slightly behind manually fine-tuned \psocake{} in terms of R-split and redundancy, it outperforms in CC$_1/2$, CC*, $I/\sigma(I)$, and completeness values.  This reinforces the effectiveness of \peaknet{} in achieving respectable results across varied sample types, without any manual intervention or parameter fine-tuning.  

Overall, \peaknet{} demonstrates comparable performance to a manually fine-tuned \psocake{} for two distinct samples, showcasing its potential for autonomous peak finding with consistent results.  By reducing dependence on the expertise of individuals, which can have considerable variation, \peaknet{} offers a promising user-independent solution.  Notably, despite primarily being trained on PK samples, \peaknet{} maintains robust and versatile performance, indicating its broad applicability and effectiveness in handling diverse datasets.

\begin{table*}
\caption{Comparison of merging statistics between \peaknet{} and fine-tuned \psocake{}.  Thaumatin (TH) results were
merged from run 31, 32, 33, 34 and 36. Proteinase (PK) results were merged from run
37 and 38.}
\label{tb : merge}
\centering
\resizebox{1.0\textwidth}{!}{
\begin{tabular}{lllllllll}
\hline
Sample              & Methods & R-split & CC$_{1/2}$     & CC*       & $I/\sigma(I)$   & Resolution range (Å) & Redundancy & Completeness \\ \hline
\multirow{2}{*}{TH} & \peaknet{} & \textbf{0.2605}  & \textbf{0.9434} & \textbf{0.9853} & \textbf{2.0590} & 1.45-27.93           & \textbf{76.05}      & \textbf{0.9594}       \\ \cline{2-9} 
                    & \psocake{} & 0.2732  & 0.9368 & 0.9835 & 2.0277 & \textbf{1.01-31.32}           & 72.20      & 0.9367       \\ \hline
                    &         &         &           &           &          &                      &            &              \\ \hline
Sample              & Methods & R-split & CC$_{1/2}$     & CC*       & $I/\sigma(I)$   & Resolution range (Å) & Redundancy & Completeness \\ \hline
\multirow{2}{*}{PK} & \peaknet{} & 0.7711  & \textbf{0.2234}    & \textbf{0.6043}    & \textbf{1.135}    & 1.45-26.98           & 16.36      & \textbf{0.8668}       \\ \cline{2-9} 
                    & \psocake{} & \textbf{0.7589}  & 0.1900    & 0.5650    & 0.974    & \textbf{1.45-30.04}           & \textbf{17.83}      & 0.8365       \\ \hline
\end{tabular}
}
\end{table*}

\subsection{Auto-masking: predicting scattering artifacts}

Auto-masking functionality in \peaknet{} represents a significant advancement in the automated processing of diffraction images.  The underlying neural network architecture enables \peaknet{} to autonomously discern and segregate pixel regions affected by scattering artifacts, such as distinct ring-like structures produced by metal shielding used to reduce air scatter and/or liquid jet nozzle scattering.  This capability circumvents the need for manual masking in the peak finding process and the need to store a binary mask used for each event which is needed for crystal indexing/integration, thereby optimizing the efficiency of crystallographic data processing.

\begin{figure*}[!ht]
\includegraphics[width=\textwidth,keepaspectratio]
{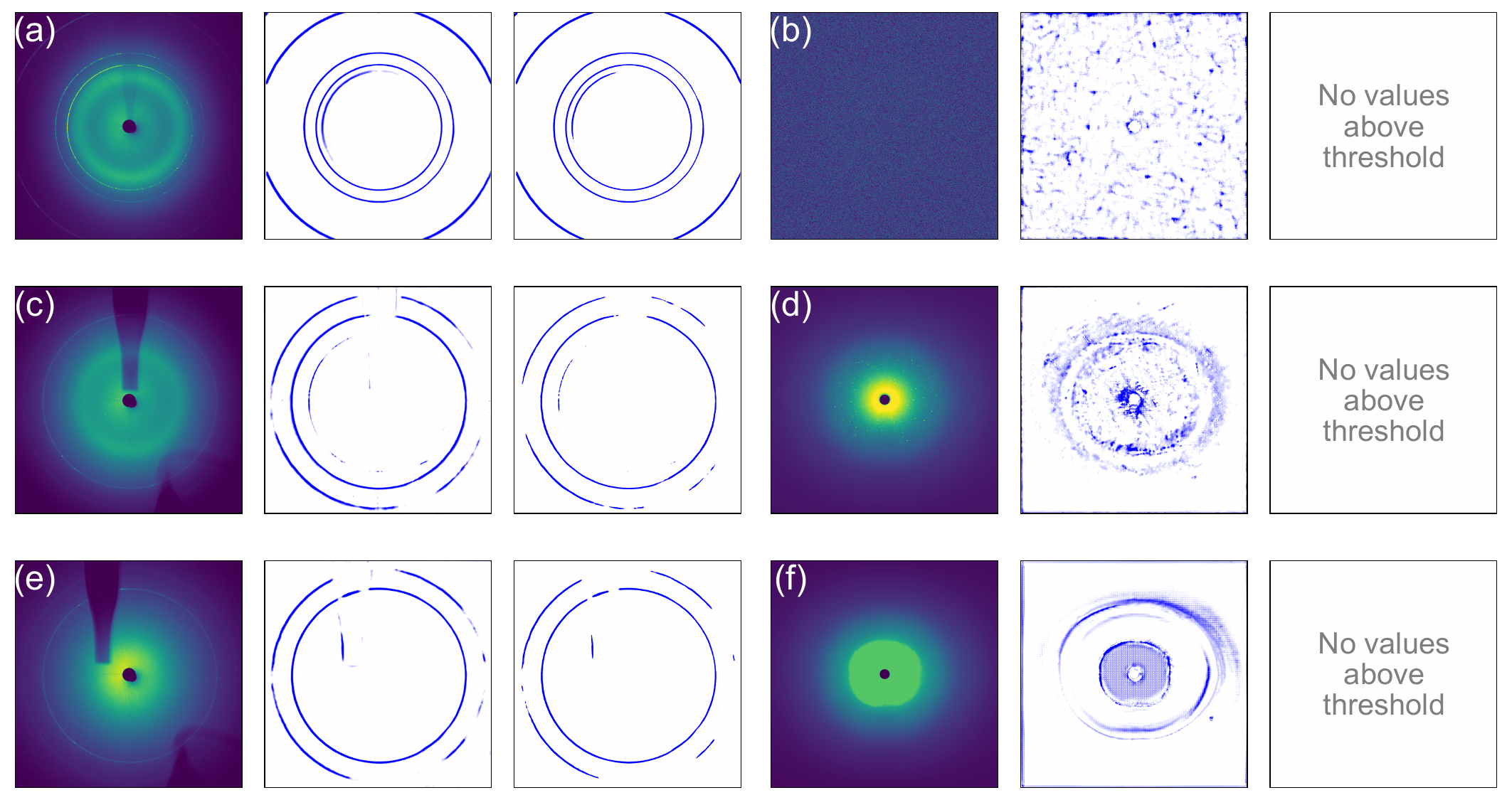} 
\caption{Examples of auto-masking using \peaknet{}.  Each panel contains three
images: the original image, the segmentation map used for masking, and the
resulting mask after thresholding.  Panels (a), (c) and (e) show three different
scenarios with artifact scattering rings.  Panels (b), (d) and (f) visualize the
response of \peaknet{} on images without any artifact scattering rings.}
\label{fig : automask 1} 
\end{figure*}

Fig. \ref{fig : automask 1} provides a comparative visualization of \peaknet{}'s effectiveness in accurately segmenting regions with bright pixels induced by scattering artifacts under various conditions.  Images without ring artifacts are also included in the visualization, reassuring that \peaknet{} does not generate or ``hallucinate" ring artifacts where they do not exist.  This visual representation serves as tangible evidence of \peaknet{} 's adeptness in discriminating and processing pertinent features, regardless of the contextual complexities.

\begin{figure*}[!ht]
\centering
\includegraphics[width=\textwidth,keepaspectratio]
{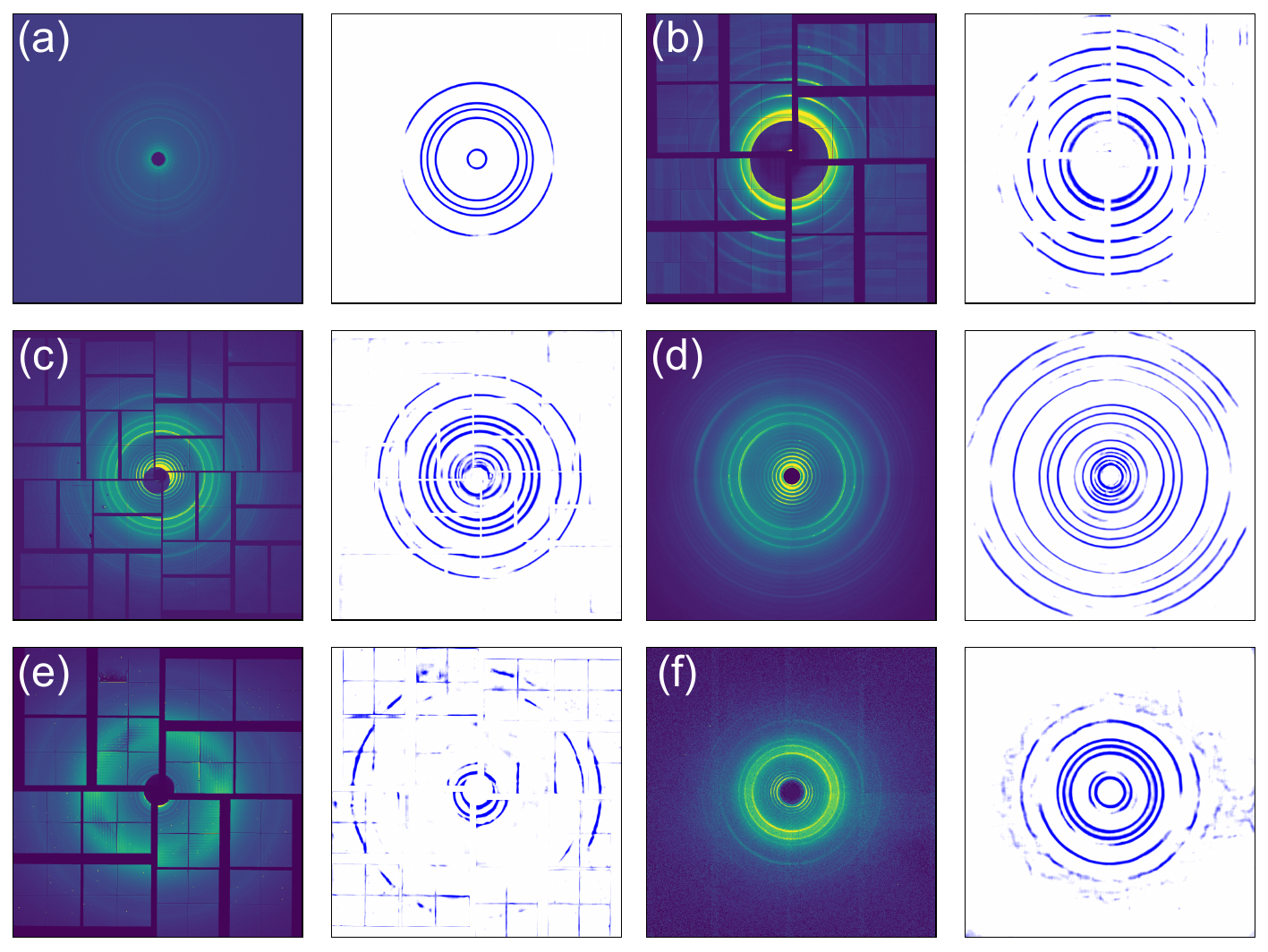}
\caption{Examples of identifying scattering rings in silver behenate images using
\peaknet{}.}
\label{fig : automask 2}
\end{figure*}

Furthermore, while evaluating \peaknet{}'s auto-masking feature on intended input images, we explored its behavior on unintended inputs, providing insights into its robustness and generalizability.  Fig. \ref{fig : automask 2} depicts illustrative examples of \peaknet{} predicting artifact scattering patterns in silver behenate scattering captured on various detectors. Silver behenate is a popular calibrant used to determine the detector position relative to the beam based on the known ring spacing. Surprisingly, \peaknet{} reliably identifies the characteristic scattering ring structures present in silver behenate images.  This observation provides compelling evidence once again, further demonstrating the robust predictive capability of \peaknet{} in accurately recognizing artifact scattering patterns.  As a result, it reinforces that effectiveness of \peaknet{}'s auto-masking feature and the potential to automatically optimize the diffraction geometry.

\subsection{Runtime efficiency}

Peak finding can pose a significant computational demand, particularly in high data rate experiments, highlighting the critical importance of an efficient peak finding algorithm.  A comparative analysis of runtime performance demonstrates the greater efficiency of \peaknet{} compared to other commonly used \psocake{} peak finder.

When processing Rayonix images with $1920 \times 1920$ pixels, \peaknet{} achieved a significantly reduced processing time of 90 milliseconds per event on an NVIDIA 1080 Ti GPU.  In contrast, \psocake{} required 176 milliseconds per event when executed on an Intel E5-2620 CPU.  Another peak finding method, RPF (robust peak finder), operating on Intel E5-2698, is reported to require 120 milliseconds but on a smaller AGIPD 1M images \citep{hadian-jaziDataReductionSerial2021}.  This result clearly demonstrates that \peaknet{} not only delivers robust Bragg peak finding results in an autonomous fashion, but also exhibits remarkable runtime efficiency.  

\begin{table*}[t]
\caption{
    The runtime performance of diffraction image analysis algorithms measured on
    Rayonix images containing 3.7 megapixels.  We measured runtime performance
    of \psocake{} and \peaknet{} directly using Rayonix images.  For reference
    purposes, we also included another peak finding method RPF
    \citep{hadian-jaziPeakfindingAlgorithmBased2017,
    hadian-jaziDataReductionSerial2021} and two classification methods for X-ray
    diffraction data reduction. RPF runtime performance was originally measured
    on AGIPD 1M \citep{allahgholiAdaptiveGainIntegrating2019}.  
}
\label{tb : runtime}
\centering
\resizebox{!}{!}{
\begin{tabular}{llcc}
\hline
Methods             & Hardware       & Image size         & Time (ms/event) \\ \hline
\peaknet{}          & NVIDIA 1080 Ti & 1920 $\times$ 1920 & 90.0            \\ \hline
\psocake{}          & Intel E5-2620  & 1920 $\times$ 1920 & 175.5           \\ \hline
RPF                 & Intel E5-2698  & 1024 $\times$ 1024 & 120.0           \\ \hline
\end{tabular}
}
\end{table*}

\section{Conclusions}

\peaknet{} exhibits exceptional capabilities in autonomous Bragg peaks finding without requiring user input or intervention, such as algorithmic parameter tuning or manual artifact masking.  The predictive performance of \peaknet{} in peak finding proves to be comparable to, if not better than, manually fine-tuned \psocake{} with user-supplied masks.  An integral feature of \peaknet{}, auto-masking, significantly enhances its ability to autonomously identify and process artifact scattering patterns.  Moreover, the efficiency of \peaknet{} is exemplified by its impressive runtime performance, surpassing commonly used peak finders like \psocake{} and RPF.  Parallel processing with GPU streaming can potentially improve the runtime efficiency even further.  In conclusion, \peaknet{}  offers a robust, versatile, and efficient solution that operates independently of user input, thereby holding significant potential for optimizing crystallographic data processing at high data rates.

\section*{Acknowledgment}

Earlier versions utilizing YOLO and U-Net models were built and tested by P.L and C.H.Y.  The experiments were designed by C.W with inputs from J.B.T and C.H.Y.  C.W implemented the residual attention U-Net, prepared the datasets, labeled experimental data, and trained and evaluated the refined neural network model.  The manuscript was written by C.W and C.H.Y with input from all authors.  This material is based upon work supported by the U.S.  Department of Energy, Office of Science, Office of Basic Energy Sciences under Award Number FWP-100643.  Use of the Linac Coherent Light Source (LCLS), SLAC National Accelerator Laboratory, is supported by the U.S.  Department of Energy, Office of Science, Office of Basic Energy Sciences under Contract No.DE-AC02-76SF00515.  C.W acknowledges the assistance provided by ChatGPT from OpenAI in refining the language and enhancing the readability of this paper.

\bibliographystyle{plainnat}
\bibliography{bibliography}

\end{document}